\RequirePackage[2020-02-02]{latexrelease}
\documentclass[10pt
,aps 
,prx
,amsmath
,amssymb
,amsfonts
,superscriptaddress
,twocolumn
,notitlepage
,longbibliography
,nofootinbib]{revtex4}
\usepackage{physics}
\usepackage[caption=false]{subfig}
\usepackage{graphicx}
\usepackage{dcolumn}
\usepackage{xcolor}
\usepackage[colorlinks=true,linkcolor=blue,urlcolor=black,citecolor=blue,bookmarksopen=true]{hyperref}
\usepackage{mathtools}
\usepackage{bm}
\usepackage{comment}

\newcommand{\CZ}{\text{CZ}} 

\begin{document}

\title{Extracting perfect GHZ states from imperfect weighted graph states via entanglement concentration}

\author{Rafail Frantzeskakis}
\affiliation{Department of Physics, University of Crete, Heraklion, 71003, Greece}

\author{Chenxu Liu}
\affiliation{Department of Physics, Virginia Tech, Blacksburg, Virginia, 24061, USA}
\affiliation{Virginia Tech Center for Quantum Information Science and Engineering, Blacksburg, VA 24061, USA}

\author{Zahra Raissi}
\affiliation{Department of Physics, Virginia Tech, Blacksburg, Virginia, 24061, USA}
\affiliation{Virginia Tech Center for Quantum Information Science and Engineering, Blacksburg, VA 24061, USA}

\author{Edwin Barnes}
\affiliation{Department of Physics, Virginia Tech, Blacksburg, Virginia, 24061, USA}
\affiliation{Virginia Tech Center for Quantum Information Science and Engineering, Blacksburg, VA 24061, USA}

\author{Sophia E. Economou}
\email{Email: economou@vt.edu}
\affiliation{Department of Physics, Virginia Tech, Blacksburg, Virginia, 24061, USA}
\affiliation{Virginia Tech Center for Quantum Information Science and Engineering, Blacksburg, VA 24061, USA}

\begin{abstract}
Photonic GHZ states serve as the central resource for a number of important applications in quantum information science, including secret sharing, sensing, and fusion-based quantum computing. The use of photon-emitter entangling gates is a promising approach to creating these states that sidesteps many of the difficulties associated with intrinsically probabilistic methods based on linear optics. However, the efficient creation of high-fidelity GHZ states of many photons remains an outstanding challenge due to both coherent and incoherent errors during the generation process. Here, we propose an entanglement concentration protocol that is capable of generating perfect GHZ states using only local gates and measurements on imperfect weighted graph states. We show that our protocol is both efficient and robust to incoherent noise errors.
\end{abstract}

\maketitle

\section{Introduction}
Photonic GHZ states play a central role in a number of quantum information applications, including quantum sensing \cite{Degen_RMP2017,lawrie2019quantum,shettell2020graph}, secret sharing \cite{Hillery_PRA1999}, and fusion-based quantum computing \cite{bartolucci2021fusionbased}. However, since photons do not interact with one another directly, creating such highly entangled states is a very nontrivial task; it in fact remains one of the most challenging problems in photonic quantum computation and communication. 

One way to generate photon-photon entanglement is through quantum interference and measurement~\cite{knill2001scheme,kok2007linear,omkar2021allphotonic}. Knill et al.~\cite{knill2001scheme} showed that it is possible to entangle photons probabilistically using only linear optical elements such as beam splitters and single-photon detectors, an approach that has been implemented in experiments~\cite{Gao2010,Li2020}. However, because the approach is probabilistic, the likelihood of successfully creating a multiphoton entangled state is exponentially small in the number of photons~\cite{knill2001scheme, bodiya2006scalable}. Although the success rate can be improved by recycling failure states and using Bell states as building blocks~\cite{browne2005resource, gimeno2015three,li2015resource}, the resource requirements of this approach continue to limit the size and fidelity of GHZ states that can be produced in this way.

A possible solution to these challenges is to employ entanglement generation methods that are deterministic, at least in principle. One such approach is to create all of the needed entanglement during the photon emission process~\cite{schon2005sequential, schon2007}. For example, Lindner and Rudolph proposed to use photon emission combined with single-qubit gates on the emitter to generate one-dimensional photonic cluster states~\cite{Lindner2009}, an approach that was subsequently demonstrated experimentally~\cite{Schwartz2016,Istrati2020}. More recently, protocols for generating more complex multiphoton entangled states, e.g., 2D cluster states~\cite{RaussendorfOneway,Economou2010,Russo2019,GSegovia2019,2021timefeedwaks}, repeater graph states~\cite{Azuma2015,Buterakos2017,hilaire2021resource}, and tree graph states~\cite{Varnava2006,Buterakos2017,Zhan_PRL2020}, have been proposed based on similar principles and experimentally demonstrated~\cite{lu2007experimental,bell2014experimental,li2019experimental,Besse2020,2021metemultinuclear}. In fact, a general algorithm for finding protocols that produce any target graph state~\cite{Hein2006} from a minimal number of quantum emitters has recently been developed~\cite{Li_npjQI2022}. However, to generate photonic graph states with low error, this approach requires nearly perfect photon emission from quantum emitters, which remains technologically challenging.

This limitation can be overcome using
a second deterministic approach to generating multiphoton entangled states. Instead of creating all the entanglement during emission, this approach leverages nonlinearities to directly implement entangling gates between photons. Taking advantage of the nonlinearity induced by strongly-coupled cavity-QED systems~\cite{Fushman2008,Loo2012}, a photon-photon controlled-Z (CZ) gate can be achieved between two incoming photons scattered by the cavity-QED system~\cite{Duan2004,nysteen2017limitations}. Given access to such CZ gates, photonic graph states can be generated on demand via time-delayed feedback~\cite{Pichler2017}.
In experiments, the phase on photonic qubits induced by the nonlinearity from the cavity-QED system is not yet fully controllable~\cite{Loo2012,kim2013quantum,Sun2016,Sun2018,Wells2019,Androvitsaneas2019}. Instead of a CZ gate, the gate that is applied to the photonic qubits is a controlled-phase (CP) gate~\cite{firstenberg2013attractive,firstenberg2016nonlinear,tiarks2016optical,thompson2017symmetry,sagona2020conditional,tiarks2019photon}. Unlike CZ gates, CP gates are not maximally entangling when the phase is not $\pi$. When the CZ gates in the graph state generation procedure are replaced by CP gates, the resulting state is called a weighted graph state~\cite{hartmann2007weighted,anders2007variational,plato2008random,hu2016universal,hayashi2019verifying}. This then leads to the question of whether or not such states are still useful; in particular, is it possible to efficiently concentrate their entanglement~\cite{bennett1996concentrating,sheng2008nonlocal,deng2012optimal, zhao2003experimental, sheng2012efficient, du2015heralded} to obtain high-fidelity GHZ states?

In this paper, we show that it is indeed possible to efficiently extract high-fidelity GHZ states from weighted graph states using only local gates and measurements. Entanglement concentration methods for imperfect graph states that maximize the entanglement in the final state have been studied extensively~\cite{paunkovic2002entanglement,zhao2003experimental,hwang2007practical,xiong2011schemes,deng2012optimal,sheng2012efficient,zhou2013efficient,choudhury2013entanglement,li2014hyperconcentration}. However, such works usually focus on generating small maximally entangled states. In our case, we propose a general protocol that can create GHZ states of any number of qubits. Our approach is also distinct from entanglement purification techniques~\cite{bennett1996mixed,bose1999purification,shi2000optimal,Dur2003,Aschauer2005,kruszynska2006entanglement,dur2007entanglement,zwerger2013universal}, which require a large number of copies of the imperfect state, as well as the application of entangling gates between copies. In contrast, our protocol uses only single-qubit measurements and single-qubit gates applied to a single copy of a 1D weighted graph state to construct a perfect GHZ state with fewer qubits. This approach can be realized in existing photonic systems with low experimental cost. 

This paper is organized as follows. In Sec.~\ref{sec:perfect}, we present our protocol for generating photonic GHZ states from 1D weighted graph states using single-qubit measurements and gates. We give the explicit example of generating a three-qubit GHZ state. 
In Sec.~\ref{sec:imperfect}, we investigate how possible experimental errors will affect our protocol. Specifically, in Sec.~\ref{subsec:coh_error}, we consider coherent errors on the CP gates, while in Sec.~\ref{subsec:incoh_error}, we focus on depolarizing errors on the weighted graph state. In Sec.~\ref{sec:summary}, we summarize our main results.

\section{Entanglement concentration with local measurements} \label{sec:perfect}
A graph state of $N$ qubits is defined based on a graph with a set of vertices ($V$) and a set of edges ($E$):
\begin{equation}
    \ket{\psi} = \prod_{\alpha,\beta \in V, \{\alpha,\beta\} \in E} \text{CZ}_{\alpha,\beta} \ket{+}^{\otimes N}, 
\end{equation}
where $\ket{+}$ is the $+1$ eigenstate of the Pauli-X operator, and the CZ gate is defined by
\begin{equation}\label{eq:def-CP}
    \CZ_{\alpha,\beta} \coloneqq \ketbra{0}{0}^{(\alpha)} \otimes I^{(\beta)} + \ketbra{1}{1}^{(\alpha)} \otimes Z^{(\beta)},
\end{equation}
where $I$ and $Z$ are the identity and Pauli-Z operator, respectively. 
In the presence of coherent errors, the $\CZ$ gate becomes a CP gate:
\begin{equation}
    \text{CP}_{\alpha,\beta} \coloneqq \ketbra{0}{0}^{(\alpha)} \otimes I^{(\beta)} + \ketbra{1}{1}^{(\alpha)} \otimes S(\phi_{\alpha,\beta})^{(\beta)},
    \label{eq:cp}
\end{equation}
where $S(\phi_{\alpha,\beta})$ is
\begin{equation}
    \text{S}(\phi_{\alpha,\beta}) = \left(\begin{array}{cc}
        1 &  \\
         & e^{i \phi_{\alpha,\beta}}
    \end{array} \right)
\end{equation}
in the computational basis with arbitrary $\phi_{\alpha,\beta}$.

\begin{figure}[h]
    \includegraphics[width = 0.4 \textwidth]{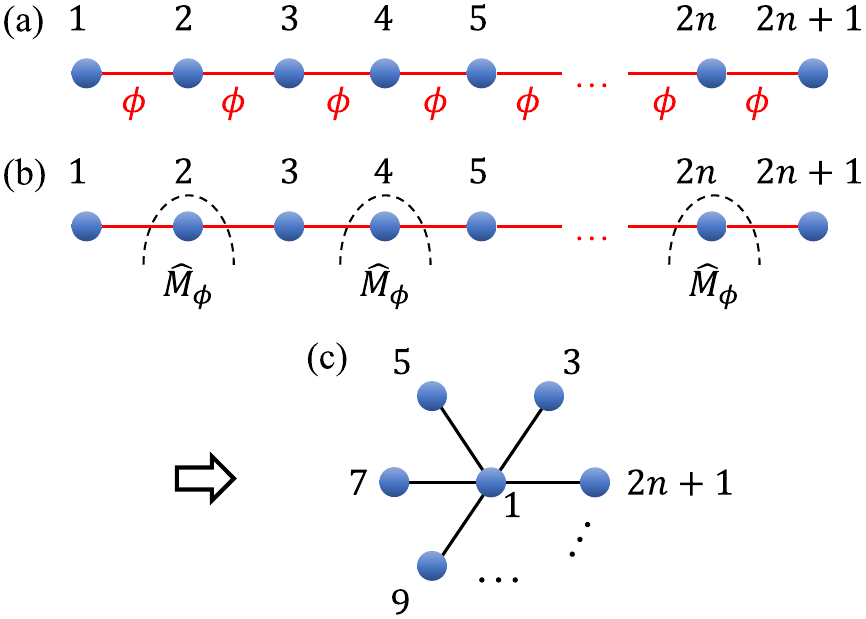}
    \caption{(a) Uniform WGS with weights $\phi$. (b),(c) The entanglement concentration protocol: (b) Start with a 1D uniform WGS containing $2n+1$ qubits and perform optimized single-qubit measurements $\hat{M}_{\phi}$ (see Eq.~\eqref{eq:measeq}) on all even qubits. 
    (c) After performing single-qubit measurements, apply a Pauli-$Z$ rotation on qubit number $2n+1$. The final graph state is a $(n+1)$-qubit GHZ state.
    }
    \label{fig:general}
\end{figure}

In the graph state generation procedure, if the CZ gates are replaced by CP gates \eqref{eq:def-CP}, the state becomes a weighted graph state (WGS). The weights of edges inside the WGS correspond to the phases of the CP gates ($\phi_{\alpha,\beta}$). In general, the weights of different edges can be different. In this paper, we focus on the case in which the edges have the same weight $\phi_{\alpha,\beta} =\phi$, and we refer to this type of WGS as a {\it uniform WGS}. 
An example of such a state is presented in Fig.~\ref{fig:general}a, where the qubits form a linear 1D uniform WGS with weight $\phi$.

Here, we consider a 1D uniform WGS with $2n+1$ qubits. 
We show that by performing single-qubit measurements on all even sites ($n$ qubits in total) and single-qubit rotations on the other qubits, the entanglement can be concentrated to probabilistically generate a $(n+1)$-qubit GHZ state, which is local unitary equivalent to a star-shaped graph state  (see Fig.~\ref{fig:general}). 
Our protocol has the following two steps:
\begin{enumerate}
    \item Measure the qubits on even sites of the 1D uniform WGS with $2n+1$ qubits (Fig.~\ref{fig:general}b) in the following basis: 
    \begin{equation}
        \hat{M}_{\phi} \coloneqq R_{z}(\phi)\hat{X} R_{z}^{\dagger}(\phi),
        \label{eq:measeq}
    \end{equation} 
    where $R_{z}(\phi) = \exp(-i \phi \hat{Z}/2)$ is a Pauli-Z rotation, 
    $\phi$ is the edge weight of the WGS, and $\hat{X}$ the Pauli-X operator. The projective measurement basis states are given in Eq.~\eqref{eq:basis}.
    The Kraus operators for each single-photon measurement are shown in Appendix~\ref{appsec:kraus}.
    \item If the measurement outcomes are all $-1$, the concentration succeeds. We then apply a Pauli-Z rotation $R_{z}[n(\pi-\phi)]$ on one of the surviving qubits to turn the final state into a $(n+1)$-qubit GHZ state. Otherwise, the concentration process failed, and we are left with a less entangled state.
\end{enumerate}

We note that the protocol succeeds only if all the measurement outcomes are $-1$.
Therefore, the success probability is
\begin{equation}
    P_{s,n}=\frac{1}{2^n} \left\vert \sin \left( \phi/2 \right)\right\vert^{2 n} \ .
    \label{eq:suc_prob}
\end{equation}

We stress that when the phase $\phi = \pi$, the uniform WGS turns into a 1D cluster state, and the measurement in~\eqref{eq:measeq} in step~1 changes into a Pauli-X measurement. In this case, there is no failure as the `failure' outcomes can be corrected by single-qubit gates on other qubits, and our protocol converts a 1D cluster state to a GHZ state deterministically. Therefore, in the rest of our paper, we focus on the case $\phi \neq \pi$. 

\begin{figure}[h]
    \includegraphics[width = 0.35 \textwidth]{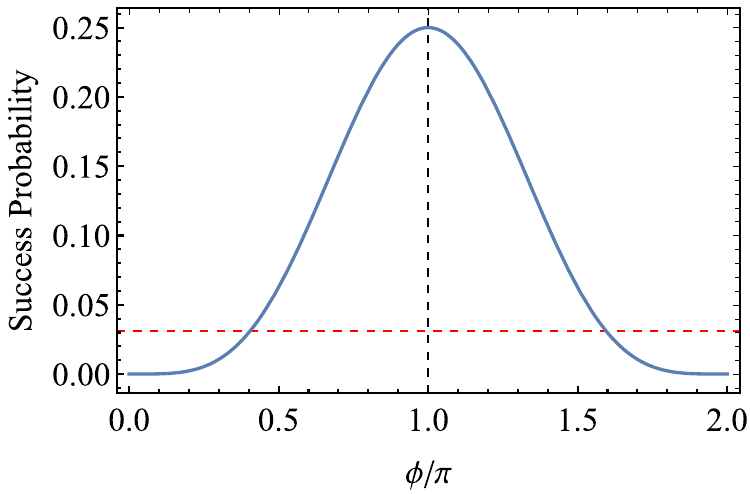}
    \caption{The success probability of constructing the 3-qubit GHZ state from a 5-qubit uniform WGS. Here, the horizontal red dashed line marks the 1/32 probability of getting a 3-qubit GHZ state using linear optical methods. 
   When $\phi = \pi$, the success probability is unity (not shown in the plot). 
   }
    \label{fig:5_to_3}
\end{figure}

In the following, we discuss an explicit example: 
We generate a $3$-qubit GHZ state from a $5$-qubit uniform WGS.
According to our protocol, we measure qubit 2 and qubit 4 in the $\hat{M}_{\phi}$ basis. When our protocol succeeds, the other three qubits are in the state
\begin{equation}
    \ket{\psi_3} = \frac{1}{\sqrt{2}} \left( \ket{000} + e^{i 2\phi} \ket{111} \right),
\end{equation}
which can be turned into a GHZ state by applying a single Pauli-Z rotation $R_{z}(\phi)$ on any of the three qubits.
In this example, the success probability is given by Eq.~\eqref{eq:suc_prob} with $n=2$, which is shown in Fig.~\ref{fig:5_to_3}.
Compared with linear optics-based methods for generating GHZ states, for which the success probability of constructing the three-qubit GHZ state starting from single photons is $1/32$ (red dashed line in Fig.~\ref{fig:5_to_3})~\cite{browne2005resource,bodiya2006scalable}, Fig.~\ref{fig:5_to_3} shows that starting from a 5-qubit uniform WGS and employing our protocol is more efficient over a wide range of weights $\phi$ ($\phi > \pi/2$). Furthermore, we stress that as long as the weights of the WGS are nonzero, there will be a non-zero probability for our protocol to succeed.

Before we move on, we comment on the implementation of the photonic CP gates in realistic experiments. Such gates are implemented either directly using linear optics or indirectly by interfering photons with quantum emitters to first create emitter-photon entanglement and then photon-photon entanglement upon measurement of the emitter. The weights $\phi_{\alpha,\beta}$ of emitter-photon CP gates reported in experiments span a wide range of values. These weights are highly sensitive to emitter-photon couplings and to detunings relative to the emitter transitions. However, accurate control of the weight has been demonstrated through interaction with a quantum dot or Rydberg atom emitter, where the resulting weights range from 0 to $\pi$ depending on the photon detuning~\cite{Loo2012,kim2013quantum,Sun2016,Sun2018,Wells2019,Androvitsaneas2019}. On the other hand, in the case of direct photon-photon gates implemented via linear optics, the weights depend on other factors. As discussed in Ref.~\cite{jeannic2021dynamical} for example, we can change the weight by sweeping the duration of the incoming pulse or by changing the time difference between the two photon pulses~\cite{firstenberg2013attractive}. The resulting CP weights range from 0 to $\sim0.55\pi$, corresponding to weakly entangled final states. Regardless of which method is used to create photon-photon entanglement,
our protocol can concentrate entanglement so long as there is entanglement shared between the photons ($\phi>0$). Moreover, our protocol enjoys a higher success probability compared to the direct, linear-optics entanglement generation approach when $\phi>\pi/2$.

\section{Noise in the protocol} \label{sec:imperfect}
In this section, we consider the robustness of our protocol to the presence of noise. As demonstrated in Refs.~\cite{Loo2012,kim2013quantum,Sun2016,Sun2018,Wells2019,Androvitsaneas2019}, the photon-photon gates implemented in experiments can achieve a large range of phases; however, precisely controlling the phase is demanding, and coherent errors in the generated WGSs are common. On the other hand, the polarization mixing of photons during their emission causes the photonic qubits to suffer from depolarization errors as well~\cite{thomas2021bright,tomm2021bright}. Therefore, in this section,
we study two types of errors: coherent errors in the weights of the given uniform WGS and incoherent errors that occur during the construction of the uniform WGS. 

In our protocol, a single measurement on one of the qubits in the WGS only affects its nearest neighbors (see Appendix~\ref{appsec:kraus}). Therefore, we at first focus on the example of constructing a two-qubit GHZ state (a Bell state) starting from a three-qubit linear WGS to understand the effect of noise on a single measurement. We find the optimal measurement basis that maximizes the entanglement between the nearest neighbors of the measured qubit to the extent possible. We show this explicitly by calculating the concurrence of the two unmeasured neighbors. Concurrence is a well-known entanglement measure for two qubits~\cite{hill1997entanglement,romero2007direct, Horodecki2009,meng2021concurrence}, but to the best of our knowledge, there does not exist a natural generalization of concurrence for $n$-qubit systems, where $n>2$. Thus for $n>2$, we instead calculate correlation functions to quantify the entanglement. In particular, to show that our entanglement concentration protocol continues to work well for larger system sizes even in the presence of noise, we calculate $ZZ$ correlation functions on multiple pairs of qubits in target GHZ states containing up to 16 qubits. We also compute the fidelity relative to a perfect GHZ state for states of up to 9 qubits.

\subsection{Coherent error} \label{subsec:coh_error}

\begin{figure*}
\includegraphics[width=\textwidth]{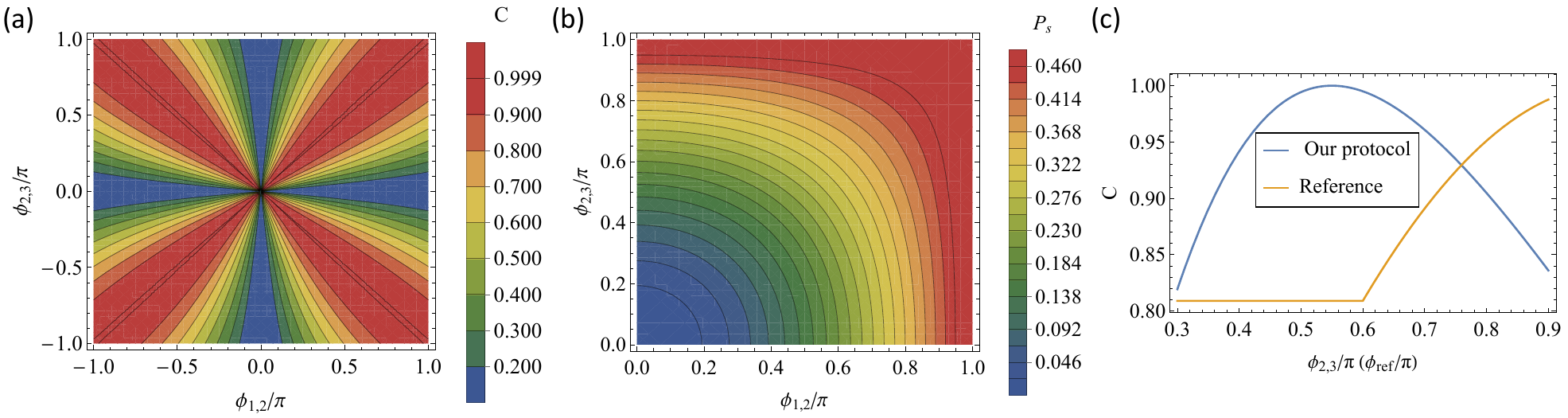}
    \caption{Constructing a 2-qubit entangled state from a 3-qubit uniform WGS with coherent error that causes the weights to differ: $\phi_{1,2} \neq \phi_{2,3}$.
     (a) The concurrence of the resulting two-qubit state when our protocol succeeds. (b) The success probability $P_s$ of our protocol. 
     (c) Concurrence (blue) of the two-qubit state after a successful measurement in our protocol in the special case of a 3-qubit WGS where $\phi_{1,2}$ is fixed to $0.55 \pi$, and $\phi_{2,3}$ can vary from $0.3\pi$ to $0.9\pi$. For comparison, we also show the concurrence of a 2-qubit WGS with weight $\phi_{\text{ref}} = \text{Max}(0.6 \pi, \phi_{2,3})$ (orange).} 
    \label{fig:coh_error}
\end{figure*}

So far, we have restricted our attention to uniform WGSs. However, due to experimental errors, the CP gates can have different phases ($\phi_{\alpha,\beta}$). The created photonic WGS may not be uniform. In this subsection, we consider the performance of our protocol when there are coherent errors on the two weights of a three-qubit WGS ($\phi_{1,2} \neq \phi_{2,3}$). We also consider the impact of errors on the concentration of larger GHZ states containing up to 16 qubits.

To check whether a GHZ state can still be prepared from a 3-qubit WGS with coherent error, we numerically optimize the concurrence of the outer two qubits after the measurement of the middle qubit by adjusting the measurement basis ($\hat{M}_\phi$) of the latter. 
The concurrence of a given two-qubit mixed state $\rho$ is defined as~\cite{Horodecki2009}
\begin{equation}
    \mathcal{C}(\rho) = \text{max}(\lambda_1-\lambda_2-\lambda_3 -\lambda_4, 0) \ ,
    \label{eq:concurrence}
\end{equation}
where $\lambda_1$, $\lambda_2$, $\lambda_3$ and $\lambda_4$ are the singular values of the matrix $\sqrt{\rho} \sqrt{\tilde{\rho}}$ with 
\begin{equation}
    \tilde{\rho} = (Y_1 \otimes Y_2) \rho^* (Y_1 \otimes Y_2)  \ .
\end{equation}
$\rho^*$ is the complex conjugate of the density matrix $\rho$ in the computational basis. We find that the concurrence is maximized when the measurement is $\hat{M}_{\phi'}$ with phase $\phi' = (\phi_{1,2} + \phi_{2,3})/2$ in Eq.~\eqref{eq:measeq}. 

Note that when $\phi_{1,2} \neq \phi_{2,3}$, the state after the measurement is not a maximally entangled state. In Fig.~\ref{fig:coh_error}a, we plot the concurrence of the two-qubit state after our protocol. When $\phi_{1,2} = \phi_{2,3}$ or $\phi_{1,2} = -\phi_{2,3}$, the concurrence can reach $1$. The latter case is consistent with a previous finding regarding entanglement concentration on WGSs with opposite-phase CP gates~\cite{kok2008effects,heo2017analysis}. 
Our protocol targets the situation in which the CP gate is close to a CZ gate but has systematic phase shifts away from a perfect CZ. 
Therefore, in Fig.~\ref{fig:coh_error}b, we consider the success probability of our protocol when $\phi_{1,2}$ and $\phi_{2,3}$ are in the range of $0$ to $\pi$. 
We can see that as the entangling power of the CP gate decreases, i.e., $\phi_{1,2}$ and $\phi_{2,3}$ decreases to $0$, the probability to get a highly entangled state decreases as well. 

In Fig.~\ref{fig:coh_error}c, we focus on the case in which the CP weights are centered at $\sim0.6\pi$ (corresponding to the approximate value of the photon-photon correlations in Ref.~\cite{jeannic2021dynamical}). We compare the concurrence of the state prepared by our protocol (blue line in Fig.~\ref
{fig:coh_error}c) with a two-qubit WGS generated by a CP gate with phase $\phi_{\text{ref}}=\text{max}(0.6 \pi, \phi_{2,3})$ (orange line in Fig.~\ref{fig:coh_error}c). We notice that there is a region, especially when $\phi_{2,3} \sim 0.55 \pi$, where our protocol improves the resulting entanglement compared to directly applying a CP gate with $\phi_{\textnormal{ref}}$. Note that even if the concurrence of the resulting state is not sufficiently high for a given application, our protocol could be combined with standard purification techniques to achieve a target value. Using our protocol to produce the initial imperfect Bell pairs could significantly reduce the number of copies and iterations needed for the purification process.

\begin{figure}
\includegraphics[width=\columnwidth]{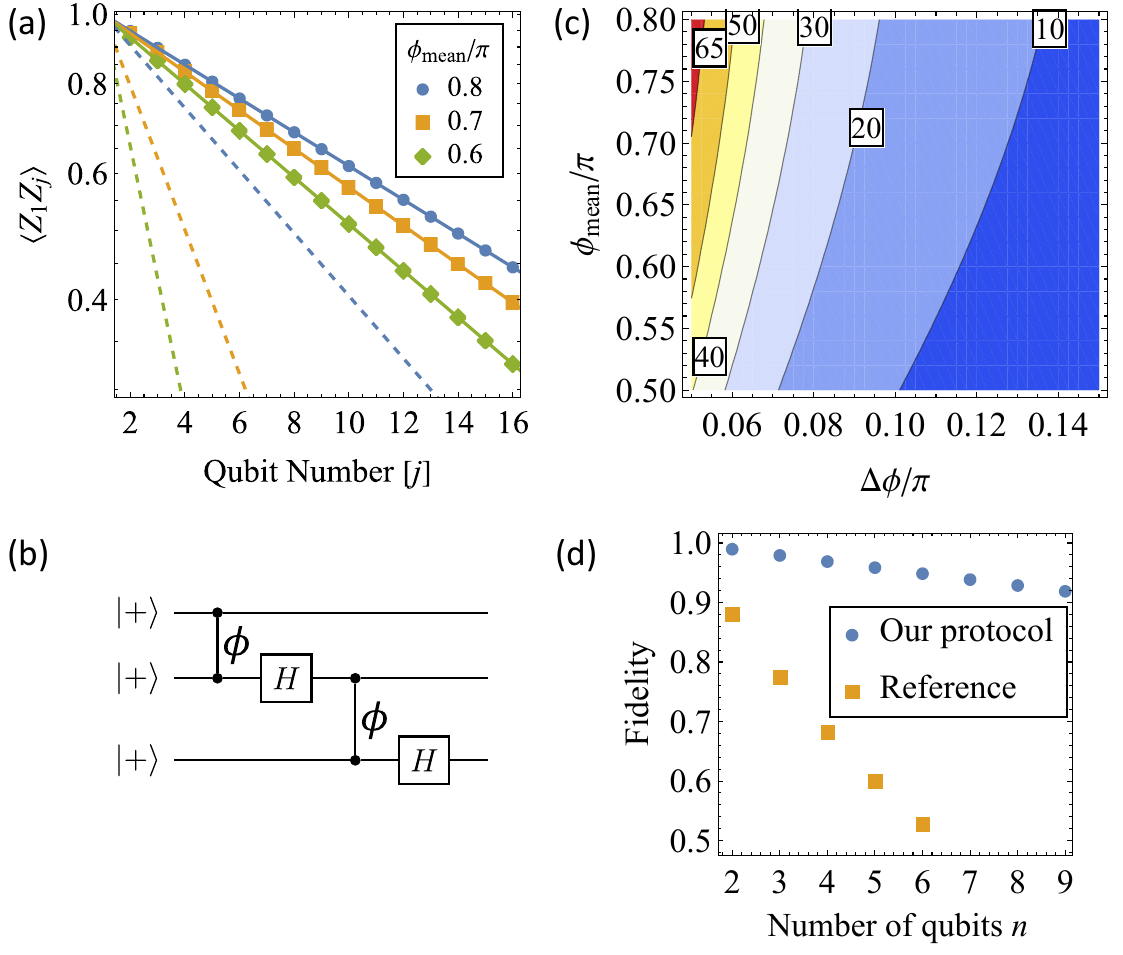}
    \caption{Effect of correlated errors on the entanglement and fidelity of concentrated GHZ states containing up to 16 qubits. (a) $ZZ$ correlation functions of 16-qubit GHZ states obtained from measuring the even qubits of 31-qubit linear WGSs with alternating CP weights $\phi_{\text{mean}} \pm \Delta \phi$ with $\Delta\phi=0.1 \pi$. $\langle Z_1 Z_j \rangle$ correlation functions with $j=2,...,16$ are shown for three different values of $\phi_{\text{mean}}$. The solid lines are fits of the data to exponential decays. For comparison, the $\langle Z_1 Z_j \rangle$ correlation functions of GHZ-like states directly generated by starting from a linear array of 16 qubits in the state $\ket{+}^{\otimes16}$ and applying CP gates with angle $\phi_{\text{mean}}$ between neighboring qubits and Hadamard gates are also shown (dashed lines). (b) 3-qubit example circuit used to create the directly generated reference states in (a) and (d). (c) The decay lengths of the correlation functions $\langle Z_1 Z_j \rangle$ with $j=2,...,16$ as a function of $\phi_{\text{mean}}$ and $\Delta \phi$ obtained from fits to exponential decays like those shown in (a). (d) Fidelity of concentrated GHZ states containing $n=2$ to $6$ qubits (blue dots) relative to a perfect GHZ state. Here, the initial linear WGS has alternating weights with $\phi_{\text{mean}}=0.55\pi$ and $\Delta\phi=0.05\pi$, which are values quoted in Ref.~\cite{jeannic2021dynamical}. For comparison, fidelities of imperfect GHZ states directly constructed using circuits as in (b) with $\phi = 0.55 \pi$ and initial states $\ket{+}^{\otimes n}$ for $n=2$ to $6$ are also shown (orange squares). Each point is obtained by applying arbitrary single-qubit gates to all qubits and adjusting gate parameters until the fidelity of the resulting state relative to a perfect GHZ state is maximized.}
    \label{fig:coh_error_scale}
\end{figure}

Next, we examine the impact of coherent errors when our protocol is scaled to larger GHZ states containing up to 16 qubits. As a concrete example, we consider starting from a linear WGS with alternating phases $\phi_{\text{mean}} \pm \Delta \phi$. Via numerical search, we find that the state fidelity relative to a perfect GHZ state is optimal when the measurement basis is chosen as in Eq.~\eqref{eq:measeq} with $\phi=\phi_{\text{mean}}$. Furthermore, we find that the single-qubit unitaries that maximize the fidelity in the case of successful measurement outcomes are always $Z$ rotations, which commute with the $ZZ$ stabilizers of the GHZ state. Therefore, to quantify how the entanglement decreases as the GHZ state grows to larger numbers of qubits, we consider the decay of the $\langle Z_1 Z_j \rangle$ correlation function, where $j$ ranges from $2$ to $n$, where $n$ is the number of qubits in the GHZ state.

Figure~\ref{fig:coh_error_scale}a shows the decay of the $\langle Z_1 Z_j \rangle$ correlation function with increasing $j$ for three different values of $\phi_{\text{mean}}$ and with $\Delta \phi = 0.1 \pi$. It is evident that the correlation function decays exponentially as the system size grows. This is because the coherent errors on the CP gates degrade the projective measurements on the even qubits of the WGS, which gradually decrease the long-range entanglement in the resulting GHZ-like state. Note that while all qubits can be viewed as nearest-neighbors in a perfect GHZ state, the fact that here we start from a noisy, linear WGS produces an asymmetry and effective spatial ordering of the qubits in the final GHZ-like state. We fit the correlation function and extract its decay length. For comparison, results for the direct construction of GHZ-like states are shown as well. In particular, the figure shows $\langle Z_1 Z_j \rangle$ correlation functions of GHZ-like states generated by starting from a linear array of qubits in the state $\ket{+}^{\otimes16}$ and sequentially applying CP gates with angle $\phi_{\text{mean}}$ between all pairs of neighboring qubits followed by Hadamard gates on one qubit in each pair (see Fig.~\ref{fig:coh_error_scale}b for a 3-qubit example circuit). It is evident from the figure that the correlation functions decay much more quickly in this case compared to using our entanglement concentration protocol.

In Fig.~\ref{fig:coh_error_scale}c, we plot the decay lengths extracted from fits like those shown in Fig.~\ref{fig:coh_error_scale}a as a function of $\phi_{\text{mean}}$ and $\Delta \phi$. When the mean phase $\phi_{\text{mean}}$ is close to $\pi$, the decay length decreases more slowly as $\Delta \phi$ increases.

To provide further evidence that our protocol can generate better GHZ states compared to direct generation, even when coherent errors on the CP gates are included, we also compute the fidelity relative to a perfect state starting from a linear WGS with alternating weights, with $\phi_{\text{mean}} = 0.55\pi$ and $\Delta\phi= 0.05\pi$, which is shown in Fig.~\ref{fig:coh_error_scale}d (blue dots). For comparison, we also show fidelities for directly generated GHZ-like states constructed using circuits like that shown in Fig.~\ref{fig:coh_error_scale}b. In each case we optimize the fidelity relative to a perfect GHZ state by performing arbitrary single-qubit gates on all qubits and adjusting gate parameters. The optimal fidelity is shown in Fig.~\ref{fig:coh_error_scale}d (orange squares). Even with coherent errors, our protocol can generate GHZ states with better fidelity, especially as the size of the target state grows.

\subsection{Photon depolarization error} \label{subsec:incoh_error}

\begin{figure*}
\includegraphics[width=\textwidth]{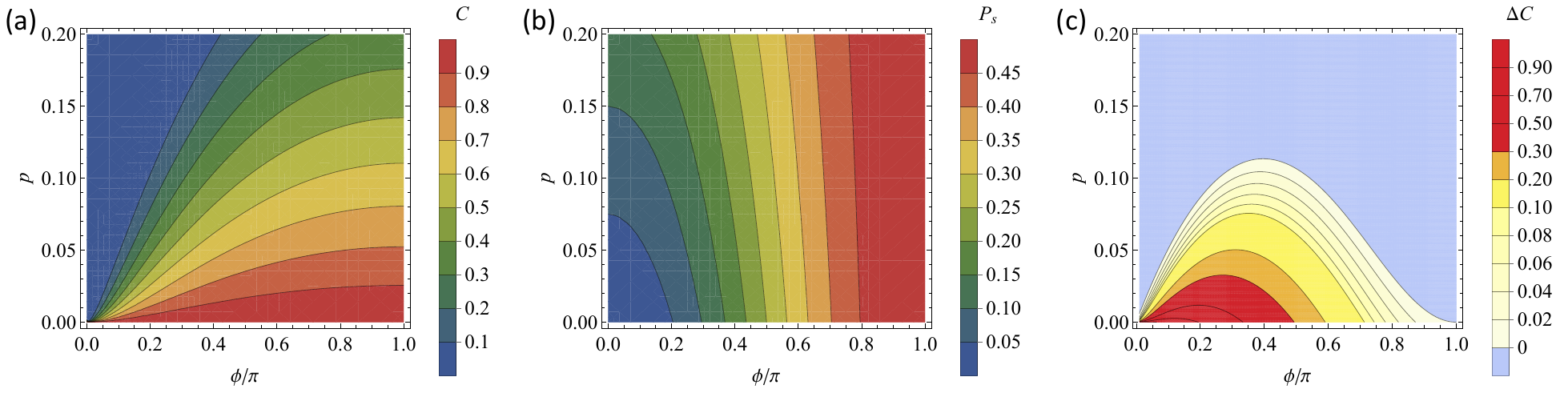}
    \caption{The performance of our protocol in the presence of depolarizing errors on the initial linear WGS.
    (a) The concurrence $C$ of the 2-qubit state after successfully applying our protocol on a 3-qubit WGS with depolarizing error. The concurrence is shown as a function of CP weight $\phi$ and depolarizing error probability $p$. (b) The success probability of our protocol as a function of $\phi$ and $p$. (c) Comparison of the concurrences of the state from our protocol with that of a 2-qubit uniform WGS in the presence of the same depolarizing error. The concurrence advantage ($\Delta C$) is shown as a function of $\phi$ and $p$.}
    \label{fig:incoh_error}
\end{figure*}

Here, we study the performance of our protocol by considering possible incoherent errors on the initial linear, uniform WGS. Due to photon scattering and frequency fluctuations of quantum emitters, the photonic qubits are likely to suffer from dephasing and depolarization errors during the graph state generation process. The depolarizing error for a single qubit is described by the channel~\cite{nielsen2002quantum}
\begin{equation}
    \mathcal{E}(\rho) = (1- p )\rho +\frac{p}{3}\left(X \rho X + Y \rho Y + Z \rho Z\right) \ ,
    \label{eq:dep_error}
\end{equation}
with error probability $p$, while a dephasing error can be modeled by a single Pauli-$Z$ error on the qubit, such that the impact of dephasing errors is contained in the analysis of the depolarizing error. 
Therefore, here we focus on the case of depolarizing error and leave the discussion of dephasing errors to Appendix~\ref{appsec:dephase}. In this section, we investigate the impact of such errors on the concentration of GHZ-like states involving up to 12 qubits.

To incorporate depolarizing errors into our calculations, we apply the same depolarizing error model to all the photonic qubits before applying the CP gates that prepare the initial linear WGS. This is because we assume the errors are equally likely to happen on all photonic qubits, which can be though of as the worst-case scenario.

To gain an understanding of how depolarizing errors affect the measurement basis in our protocol, we first consider the simple case of a three-qubit linear WGS in which the middle qubit is measured to produce an approximate Bell state. We numerically optimize the concurrence of the two unmeasured qubits by adjusting the measurement basis, as the concurrence (Eq.~\eqref{eq:concurrence}) is still a good measure of entanglement for two-qubit mixed states. Our numerical analysis reveals that the optimal measurement basis is not affected by the depolarizing error. However, the two-qubit state is mixed and no longer a maximally entangled state regardless of the CP angle $\phi \in (0, \pi)$.

To further understand how depolarizing error affects our protocol, in Fig.~\ref{fig:incoh_error}a, we show the best concurrence of the final two-qubit state after applying our protocol. We sweep the uniform WGS weight $\phi$ and the depolarizing error probability $p$ on the photonic qubits. Notice that even with moderate depolarizing error ($p<0.02$), our protocol can still generate a decent amount of entanglement ($C > 0.9$) over a wide range of weights $\phi$. 
In Fig.~\ref{fig:incoh_error}b, we show the success probability of our protocol. In the range $\phi > 0.6 \pi$, the success probability is weakly dependent on the depolarizing error probability $p$. Our protocol is more robust against depolarizing error in the regime where $\phi$ is close to $\pi$. As $\phi$ decreases, i.e., the CP gate generates less and less entanglement directly, the entanglement generated by our protocol drops significantly with stronger depolarizing error [see Fig.~\ref{fig:incoh_error}a, $\phi \sim 0.1 \pi$ for example].
This is because the symmetry of the depolarization process does not prefer a special measurement basis. However, the depolarizing error makes the final state mixed, which decreases the entanglement of the state. 

\begin{figure}
\includegraphics[width=\columnwidth]{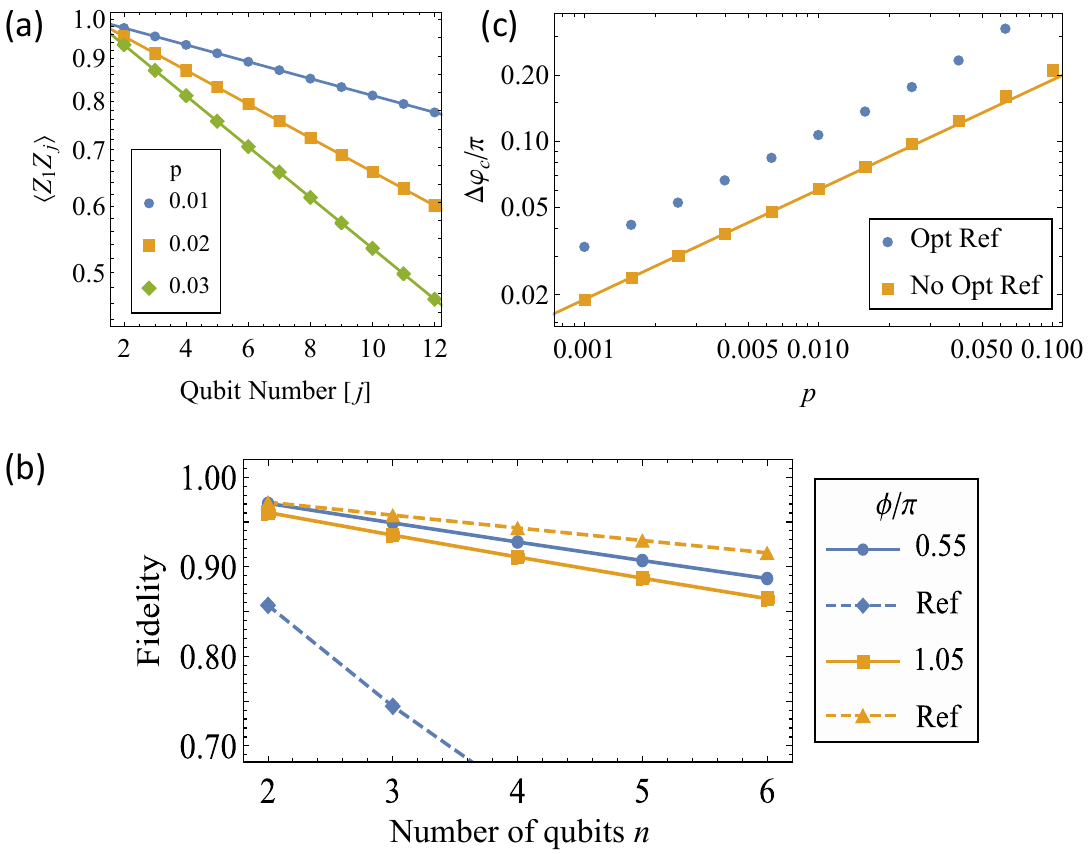}
    \caption{Effect of depolarization errors on the entanglement and fidelity of concentrated GHZ states containing up to 12 qubits. (a) $ZZ$ correlation functions of 12-qubit GHZ states obtained from measuring the even qubits of 25-qubit linear WGSs with uniform CP weights $\phi=0.55\pi$. $\langle Z_1 Z_j \rangle$ correlation functions with $j=2,...,12$ are shown for three different values of the depolarization probability $p$. The solid lines correspond to fits of the data to exponential decays. (b) Fidelity of concentrated GHZ states containing $n=2$ to $6$ qubits relative to a perfect GHZ state. Here, the initial linear WGS has the same depolarization probability $p=0.02$ on all photons. Results are shown for two different values for the CP gates: $\phi=0.55\pi$~\cite{jeannic2021dynamical} (blue solid lines) and $1.05\pi$~\cite{tiarks2016optical} (orange solid lines).  For comparison, fidelities of imperfect GHZ state directly constructed by sequential applying CP gates with the same CP gates are also shown (dashed lines). The generation circuit is shown in Fig.~\ref{fig:coh_error_scale}d, where the initial states of photonic qubits experience the same depolarization error ($p=0.02$). (c) The critical CP angle ($\Delta\varphi_c \equiv \pi - \phi_c$) at which our protocol and the direct (reference) approach have the same fidelity as a function of the depolarizaton error probability $p$ in the case of $n=2$ qubits. Results with (blue dots) and without (orange squares) optimization of the reference state with respect to single-qubit gates are shown. The orange line is an analytical result obtained using perturbation theory.
    }
    \label{fig:incoh_error_scaling}
\end{figure}

To see whether and to what extent we can still benefit from using our protocol, we calculate the concurrence advantage ($\Delta C$) of our protocol in Fig.~\ref{fig:incoh_error}c. We define the concurrence advantage as the  difference between the concurrence of the two-qubit state generated by our protocol and the reference concurrence, which is the concurrence of a two-qubit state generated by directly applying a CP gate with phase $\phi$ on a pair of depolarized photonic qubits with the same error probability.
We notice that over a large range of parameters (the region with warm colors in Fig.~\ref{fig:incoh_error}c), our protocol will result in more entanglement between the two photonic qubits. Specifically, in the region where $\phi \sim 0.8 \pi$, our protocol can generate more entanglement even with $p \sim 5\%$ [Eq.~\eqref{eq:dep_error}]. Although the two-qubit state is mixed and does not have maximal entanglement, the state with more entanglement can help to increase the efficiency with further entanglement purification.

To investigate the effect of depolarizing errors on concentrating GHZ states with more photons, we calculate $ZZ$ correlation functions of the outcome state of our protocol as in the case of coherent errors discussed above. The results for a 12-qubit GHZ-like state concentrated from an initial 25-qubit linear WGS are shown in Fig.~\ref{fig:incoh_error_scaling}a for three different values of the error probability $p$ and a CP angle $\phi=0.55\pi$. We note that when $p=0$, our protocol can generate perfect GHZ states, which gives a constant correlation function $\langle Z_1Z_j\rangle=1$ for all $j$. As we increase the depolarizing error probability $p$, the correlations start to decay exponentially with a characteristic decay length. This is similar to the discussion in Sec.~\ref{subsec:coh_error}. Note that as shown in Refs.~\cite{thomas2021bright,tomm2021bright}, there is no possibility of creating pure polarized photons, and we instead have mixed polarization states. In quantum dot experiments, the polarization mixing is around 2-5\%. We therefore consider a depolarization error probability of $p=0.02$ as a concrete example. In Fig.~\ref{fig:incoh_error_scaling}b, we calculate the fidelity of the state generated from our protocol with CP weight $\phi=0.55\pi$ (blue dots). For comparison, we also calculate the state fidelity (optimized over local gates) generated directly using the same CP gates and with the same depolarization error $p=0.02$, which is shown as the blue diamonds on a dashed line in Fig.~\ref{fig:incoh_error_scaling}b. It is evident that our protocol yields substantially better GHZ states even in the presence of realistic photon polarization errors.

On the other hand, Fig.~\ref{fig:incoh_error_scaling}b also reveals that when the CP weight is $1.05 \pi$ as in Ref.~\cite{tiarks2016optical}, the reference state has better fidelity. This is because the main source of error in the state produced from our entanglement concentration protocol is the depolarizing error, while for the reference state, the imperfection in the CP gate plays a more important role. The relative performance of our protocol versus direct state generation thus depends on the relative importance of $p$ and $\phi-\pi$. For a given value of the depolarizing error probability $p$, there is in fact a critical value $\phi_c$ of the CP weight at which the fidelities of the two state-generation methods exactly match. Our protocol does not provide any benefit when $\phi>\phi_c$. Figure~\ref{fig:incoh_error_scaling}c shows the dependence of $\phi_c$ on $p$ in the case of $n=2$ qubits in the final state, from which it is evident that our protocol provides a benefit across a range of CP weights that shrinks from $\phi<0.98\pi$ to $\phi<0.8\pi$ as $p$ increases by two orders of magnitude from 0.001 to 0.1. Further details about these results can be found in Appendix~\ref{appsec:critical_phase}.

\section{Conclusions} \label{sec:summary}

In this paper, we addressed the problem of concentrating the entanglement of 1D weighted graph states through single-qubit measurements. We proposed a protocol to probabilistically construct GHZ states (or equivalently, star-shape graph states) using 1D uniformly weighted graph states.
The protocol only uses single-qubit measurements and gates, which can be applied efficiently in photonic systems. We showed that the protocol can efficiently generate small-sized GHZ states with a large tolerance on WGS weights (the controlled phase gate angles) compared to the generation of GHZ states using linear optical methods.
Our protocol can generate more entanglement compared to other approaches in the presence of moderate coherent two-qubit or single-qubit depolarizing errors on the photonic qubits.

\section{Acknowledgments}

We thank Paul Hilaire for the fruitful discussions. S.E.E. acknowledges support from the NSF (grant no.
1741656), the EU Horizon 2020 programme (GA 862035 QLUSTER), and the Commonwealth Cyber Initiative (CCI). E.B. acknowledges support from NSF grant no. 2137953.

\appendix

\begin{figure*}
\includegraphics[width=\textwidth]{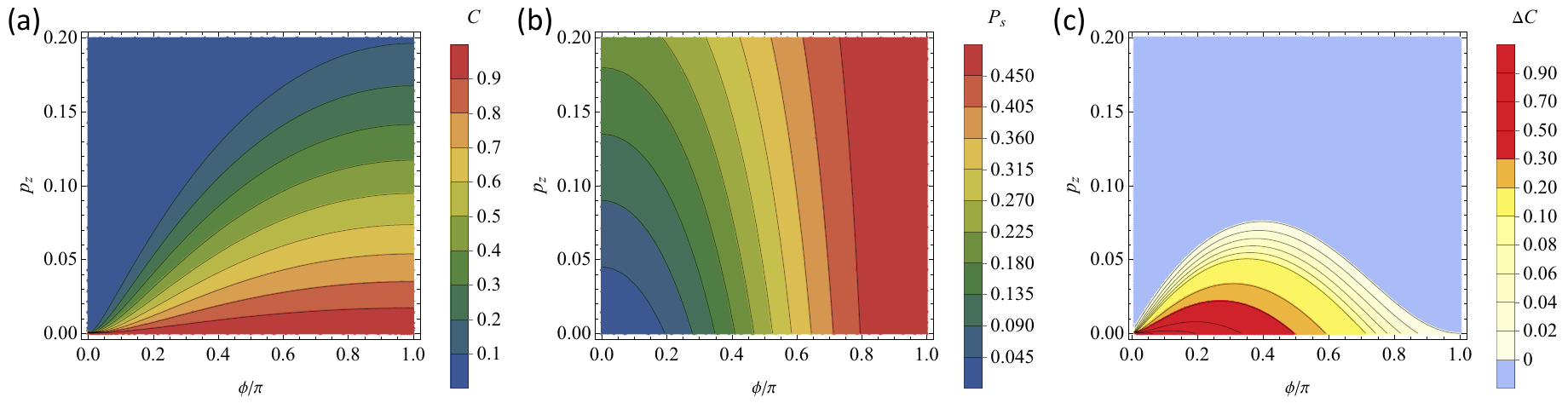}
    \caption{The performance of our protocol in the presence of dephasing errors on the initial linear WGSs.
    (a) The concurrence $C$ of the 2-qubit state after successfully applying our protocol on a 3-qubit WGS with dephasing error. The concurrence is shown as a function of CP weight $\phi$ and dephasing error probability $p_z$. (b) The success probability of our protocol as a function of $\phi$ and $p_z$. (c) Comparison of concurrences of the state from our protocol with that of a 2-qubit uniform WGS in the presence of the same dephasing error $p_z$. The concurrence advantage ($\Delta C$) is shown as a function of $p_z$ and $\phi$.}
    \label{fig:dephasing_error}
\end{figure*}   
\section{Kraus operators}\label{appsec:kraus}

In this section, we present the Kraus operators of our entanglement concentration protocol. The elementary operation in our protocol is the single-qubit measurement $\hat{M}_\phi$ on the middle qubit in a 3-qubit uniform linear WGS. 

As we discussed in the main text, a 3-qubit uniform WGS can be generated by applying CP gates between neighboring qubits:
\begin{equation}
\ket{\psi} = \text{CP}_{1,2} \text{CP}_{2,3} \ket{+++}.
\end{equation} 
In our protocol, we perform a single-qubit measurement on qubit 2, 
with $\hat{M}_{\phi}$ as shown in Eq.~\eqref{eq:measeq}, which is a projective measurement in the basis
\begin{equation}
	 \ket{\pm_\phi} = \frac{1}{\sqrt{2}} \left( \ket{0} \pm e^{-i \phi} \ket{1} \right). 
	 \label{eq:basis}
\end{equation}
Therefore, the Kraus operators on qubits 1 and 3 corresponding to $\pm 1$ measurement outcomes are
\begin{align}
	K_{\pm} = \bra{\pm_\phi}_{2} \text{CP}_{1,2} \text{CP}_{2,3} \ket{+}_{2}.
\end{align}
Using the expression for the CP gates [Eq.~\eqref{eq:cp}], the Kraus operators can be expressed as
\begin{align}
	K_{+} & = \cos \left(\frac{\phi}{2} \right) \left( e^{-i\phi/2} \dyad{0 0} + e^{i\phi/2} \dyad{1 1} \right) \nonumber \\
	& \quad + \left( \dyad{0 1} + \dyad{1 0} \right), \\
	K_{-} & = i \sin \left(\frac{\phi}{2} \right)
	 \left(e^{-i\phi/2} \dyad{0 0} - e^{i\phi/2} \dyad{1 1} \right).
\end{align}
When our protocol succeeds, i.e., the measurement result is $-1$, the measurement effectively projects the state of qubits 1 and 3 into the even parity subspace, which creates a GHZ state with probability $\sin^2(\phi/2)/2$.

\section{Photon dephasing error} \label{appsec:dephase}

In the main text, we study the effect of depolarizing errors on our protocol. Apart from depolarizing errors, dephasing error also frequently arises in photonic systems. 
Therefore, in this appendix, we consider the effect of dephasing errors, where the dephasing error for a single qubit is modeled by~\cite{nielsen2002quantum}:
\begin{equation}
    \mathcal{E}(\rho) = (1- p_z )\rho +p_z Z \rho Z,
    \label{eq:deph_error}
\end{equation}
where $p_z$ is the error probability, and $\rho$ is the density matrix corresponding to the qubit that experiences the dephasing error.

We apply dephasing errors on photonic qubits before the CP gate that generates the uniform WGS, just like we do for depolarization errors.
Similar to the argument in the main text, dephasing errors are likely to happen on all three photonic qubits in the WGS, which is used for our protocol to generate a two-qubit maximally entangled state. So, we apply the same dephasing error to all three photonic qubits, i.e., we use the same $p_z$ for all three qubits.

With the dephasing errors, we numerically optimize the measurement basis in our protocol to maximize the concurrence of the two-qubit state. We find that the optimal measurement basis is unchanged. In Fig.~\ref{fig:dephasing_error}a, we show the concurrence of the resulting two-qubit state while sweeping the three-qubit WGS weights $\phi$ and the strength of dephasing errors $p_z$. Our protocol can still achieve relatively high entanglement ($C > 0.9$) with moderate dephasing error $p_z < 0.02$. When the dephasing error strength increases, the entanglement between the two remaining, unmeasured qubits decreases. In Fig.~\ref{fig:dephasing_error}b, we show the success probability of our protocol. 

In order to understand if our protocol can generate better entanglement compared to directly generating entanglement using CP gates between photon qubits with dephasing error, we calculate the entanglement advantage $\Delta C = C - C_{\text{ref}}$, where $C$ is  the concurrence from our protocol, and $C_{\text{ref}}$ is the concurrence of the directly generated photonic states with CP gates with the same phase, which is shown in Fig.~\ref{fig:dephasing_error}c. We see that with moderate dephasing errors, our protocol can still generate more entanglement.

\section{Critical CP weights in presence of depolarization errors} \label{appsec:critical_phase}

In the main text, we show that with depolarization error $p = 0.02$, and CP weight $\phi = 1.05 \pi$, the GHZ states generated by our protocol do not have higher fidelity compared with the GHZ state directly generated by the imperfect CP gates (see Fig.~\ref{fig:incoh_error_scaling}c). In this section, we investigate when our protocol can perform better and why our protocol can sometimes fail to provide an improvement. For illustrative purposes, we consider a 3-qubit uniform WGS from which we concentrate a two-qubit GHZ state.

Note that for the reference state, we construct an imperfect GHZ state directly using the CP gates according to the gate sequence shown in Fig.~\ref{fig:coh_error_scale}b. After applying the gate sequence, we numerically optimize the state fidelity by adjusting the parameters of single-qubit gates applied to all the qubits. We stress that even without the single-qubit gates, the reference state can still achieve better fidelities when the CP gate is close to $\pi$.


\begin{figure}
    \includegraphics[width =  0.8\columnwidth ]{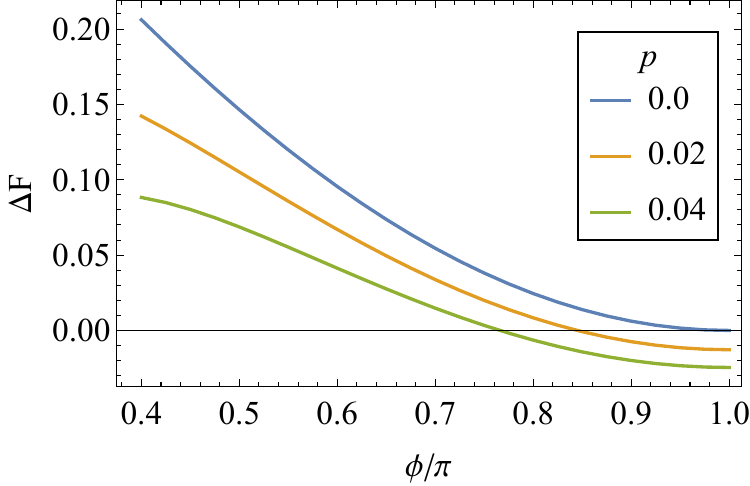}
    \caption{The difference in fidelity $\Delta F$ of our entanglement concentration protocol minus that of the reference state in the presence of photon depolarization error of probability $p$ in the case where the target state is a 2-qubit GHZ state. $\Delta F$ is shown as a function of the CP gate angle $\phi$.}
    \label{fig:dfdepolarization}
\end{figure}

In our protocol, as discussed in the main text, when there is no depolarization error on the measurement qubit, a ``successful" measurement outcome results in a perfect parity projection on the other qubits (see Appendix~\ref{appsec:kraus}) regardless of the CP gate angle. However, when there are depolarization errors, the operation is imperfect. This imperfection will be mainly affected by the depolarization error and weakly depend on the CP gates. On the other hand, the main imperfection in the reference state comes from the CP gate between the two qubits. Therefore, we expect there to be a critical CP gate angle beyond which our protocol performs worse than the reference case.

In Fig.~\ref{fig:dfdepolarization}, we plot the difference in fidelity between our protocol and the reference case. Similar to the results shown in the main text, in the reference case, we apply single-qubit gates on both qubits to numerically optimize the state fidelity. We see that, depending on the value of the depolarization error probability $p$, there is a critical value $\phi_c$ of the CP weight above which our protocol performs worse than the direct-generation approach.

To obtain a quantitative understanding of the relation between $\phi_c$ and $p$, we compute the fidelities analytically using perturbation theory. We first note that for our concentration protocol, we can apply a single-qubit gate $U$ on one of the two un-measured qubits to maximize the state fidelity to a perfect GHZ state, where 
\begin{equation}
    U = \left( 
    \begin{array}{cc}
        e^{-i(\pi-\phi)} &  \\
         & 1
    \end{array}
    \right).
\end{equation}
The optimal state fidelity is 
\begin{equation}
    F = \left( \frac{3}{3-2p} + \frac{24p}{9+(3 \cos \phi+4p)(4p-3)} \right)^{-1},
    \label{eq:our_fidelity_depolar}
\end{equation}
where $p$ is the depolarization error probability shown in Eq.~\eqref{eq:dep_error}, and $\phi$ is the CP gate angle. 

In the reference case, as it is hard to find an analytical expression for the numerically optimized single-qubit gate parameters, we focus instead on the state directly generated using the circuit shown in Fig.~\ref{fig:coh_error_scale}b (without single-qubit gates after the circuit). The fidelity relative to a two-qubit GHZ state with photon depolarization errors is
\begin{equation}
    F_{\text{no-opt}} = \frac{1}{72}\left[18 + (4p-3)(4p-9)(1-\cos\phi)\right].
    \label{eq:no-opt_ref_fidelity}
\end{equation}
Note that in the reference case, when $\phi \rightarrow \pi$, the fidelity is
\begin{equation}
    F_{\text{no-opt}} = \frac{1}{9} (9 - 12 p + 4 p^2) \sim 1-\frac{4}{3} p,
\end{equation}
which is caused by the depolarization error on two photonic qubits. However, in our protocol, when the CP gate is close to $\pi$, and $p$ is small, the state fidelity is
\begin{equation}
    F = \frac{1}{27} (3 - 2 p) (9 - 12 p + 8 p^2) \sim 1 - 2p,
\end{equation}
which suffers more from the depolarization error on the measured qubits. 

In Fig.~\ref{fig:incoh_error_scaling}c, we plot the critical phase ($\phi_c$) above which our protocol cannot improve the state fidelity. Specifically, the critical phase $\phi_c$ corresponding to un-optimized reference states is plotted as orange squares. With Eq.~\eqref{eq:our_fidelity_depolar} and Eq.~\eqref{eq:no-opt_ref_fidelity}, we can further expand the expression $F - F_{\text{no-opt}}$ in the regime where $p \rightarrow 0$ and $\phi \rightarrow \pi$ and solve for the critical phase $\phi_c$:
\begin{equation}
    \Delta \varphi_c ^2 \sim \frac{32}{9} p,
\end{equation}
where $\Delta \varphi_c \equiv \pi - \phi_c$. This result is shown as the orange line in Fig.~\ref{fig:incoh_error_scaling}c, which matches well with the numerical solution. For comparison, we also calculate the critical phase when the reference state fidelity is optimized using single-qubit gates (blue dots in Fig.~\ref{fig:incoh_error_scaling}c). We notice that the parametric dependence is the same relative to the un-optimized reference case, and the parameter dependence from our analysis still applies.

\bibliography{ref2}
    
\end{document}